\documentclass{article}

\usepackage[final, nonatbib]{neurips_2019_ml4ps}

\usepackage[utf8]{inputenc} % allow utf-8 input
\usepackage[T1]{fontenc}    % use 8-bit T1 fonts
\usepackage{hyperref}       % hyperlinks
\usepackage{url}            % simple URL typesetting
\usepackage{booktabs}       % professional-quality tables
\usepackage{amsfonts}       % blackboard math symbols
\usepackage{nicefrac}       % compact symbols for 1/2, etc.
\usepackage{microtype}      % microtypography
\usepackage{graphicx}
\title{Hunting for Dark Matter Subhalos in Strong Gravitational Lensing with Neural Networks}

\author{
  Joshua Yao-Yu Lin\thanks{equal contribution} \\
  University of Illinois at Urbana-Champaign\\
  Urbana, IL 61801 \\
  \texttt{yaoyuyl2@illinois.edu} \\
   \And
   Hang Yu$^*$ \\
   University of Illinois at Urbana-Champaign \\
   \texttt{hangyu5@illinois.edu} \\
   \AND
   Warren Morningstar \\
   Stanford University \\
  \texttt{wmorning@stanford.edu} \\
   \And
   Jian Peng \\
   University of Illinois at Urbana-Champaign \\
   \texttt{jianpeng@illinois.edu} \\
   \And
   Gilbert Holder \\
   University of Illinois at Urbana-Champaign \\
   \texttt{gholder@illinois.edu} \\
}

\begin{document}

\maketitle

\begin{abstract}
Dark matter substructures are interesting since they can reveal the properties of dark matter. Collisionless N-body simulations of cold dark matter show more substructures compared with the population of dwarf galaxy satellites observed in our local group. Therefore, understanding the population and property of subhalos at cosmological scale would be an interesting test for cold dark matter. In recent years, it has become possible to detect individual dark matter subhalos near images of strongly lensed extended background galaxies. In this work, we 
discuss the possibility of using deep neural networks to detect dark matter subhalos, and showing some preliminary result with simulated data. We found that neural networks not only show promising results on detecting multiple dark matter subhalos, but also learn to reject the subhalos on the lensing arc of a smooth lens where there is no subhalo.
\end{abstract}

\section{Introduction}

In cosmology, the cold dark matter (CDM) model, serves as the standard paradigm for cosmological research in the past decade. Despite their success on large scale, there exist several tensions on small-scale (sub-galactic) structures. The collisionless N-body simulations of cold dark matter show more substructures compared with the population of dwarf galaxy satellites observed in our local group \cite{de2010core, klypin1999missing, weinberg2015cold}. Studies indicated that baryonic physics (e.g. stellar feedback and low star-formation efficiency) that are not being modeled in the simulation could help alleviate the tensions. Others suggest that modification of the nature of dark matter (e.g. self-interaction, warm/fuzzy dark matter) could also produce a lower population of dark matter subhalos. Therefore, understanding more about the population of dark matter sub-halos would be crucial to test the cold dark matter scenario. In recent years, it has become possible to detect individual dark matter subhalos near images of strongly lensed extended background galaxies. Several groups claimed detection of dark matter subhalos via comparing smooth and perturbed lens model \cite{hezaveh2016detection, vegetti2012gravitational}, though some of them might be line-of-sight subhalos as pointed out by \cite{despali2018modelling}.

Traditional ways of detecting dark matter subhalos in strong gravitational lensing primarily based on maximum likelihood method requires precision lens modeling due to the fact that we do not know the ground truth of lens potential and the morphology of source galaxies, so one would need modeling the smooth model (strong lensing without dark matter substructures) and use penetrative method to build up a perturbed model (lensing with substructures) to search for dark matter subhalos. However, if the lensing system contains multiple subhalos, it could be quite computational expensive due to the growth of model parameters. Furthermore, we expect to detect nearly $170,000$ strongly lensed galaxies \cite{collett2015population} with LSST and Euclid in the future. It would be nearly impossible to study the details of each individual system. Therefore, automatic ways of doing data analysis would be crucial for studying dark matter substructures in the future. Machine learning, in particular deep learning would be a perfect tool for the task. 

For strong lensing science, Hezaveh et al. trained neural networks with simulated strong lensing images output of each strong lensing images correspond to five parameters of a singular isothermal ellipsoid model  \cite{hezaveh2017fast}. Their method had demonstrated that neural networks could be a powerful tool to preform fast and automated analysis on strong gravitational lensing. In this work, we further investigate the possibility of using deep neural networks to detect dark matter subhalos in strong gravitational lensing. We found that the neural networks are able to detect multiple dark matter subhalos on simulated data without further lens modeling. Furthermore, we found that just by training with the lens images as input and probability map of dark matter substructures as target, the neural networks learn to \emph{reject} subhalo on the smooth lens where there are no substructures.

\section{Strong Gravitational Lensing Simulation and the Effect of Dark Matter Subhalos}

According to General Relativity, the photons from distant galaxies would follow \emph{geodesic}. If there exist gravitational potential along the line-of-site from the observer, we would be able to see multiple images or Einstein ring made of the same source galaxy. In order to obtain a huge amount of lensing images for training, simulation of strong lensing images with subhalos is required since claimed detection of individual dark matter subhalos in strong gravitational lensing are insufficient ($\leq \mathcal{O}(10)$) for typical size of the training set for deep learning ($\approx \mathcal{O}(10^3 - 10^4)$). 

Following the lensing simulator, we use the raytracing technique to simulate strong gravitational lens images assuming thin lens approximation. In this work, we assume the subhalos are near the lens plane, so the lens could be decomposed into two parts, the smooth model described as the main halo that contributes to most of the angle of deflection, and the dark matter substructures would perturb the lensing images slightly (which is often called perturbed model). We build 2 sets of simulation, one Singular Isothermal Ellipsoid (SIE), another use a singular elliptical power law surface density profile as our macro lens models \cite{barkana1998fast}.

In this paper, we use a probability map to describe the distribution of dark matter subhalos. In simulations, we generate the lensing images with two sets of parameters ($\xi_{smooth}$ for smooth model parameters, e.g. Einstein radius $\theta_E$, ellipiticity $e1$ and $e2$, $x_{center}$ and $y_{center}$ for lens center...etc) and subhalo parameters ($\xi_{subhalo}$, e.g. $x_{i, sub}$, $y_{i, sub}$ for subhalo position, $M_{i, sub}$ for subhalo mass). Here we focus on the distribution of subhalos positions $x_{i,sub}$, $y_{i,sub}$. To generate the probability map, we define a \textit{target to density} function. It turns subhalos' positions into a Gaussian distribution with a cutoff applied. Since it is a probability problem, it would be straight to study the probability of subhalos using neural networks. 

\begin{figure}[h]
    \centering
    \includegraphics[scale=0.7]{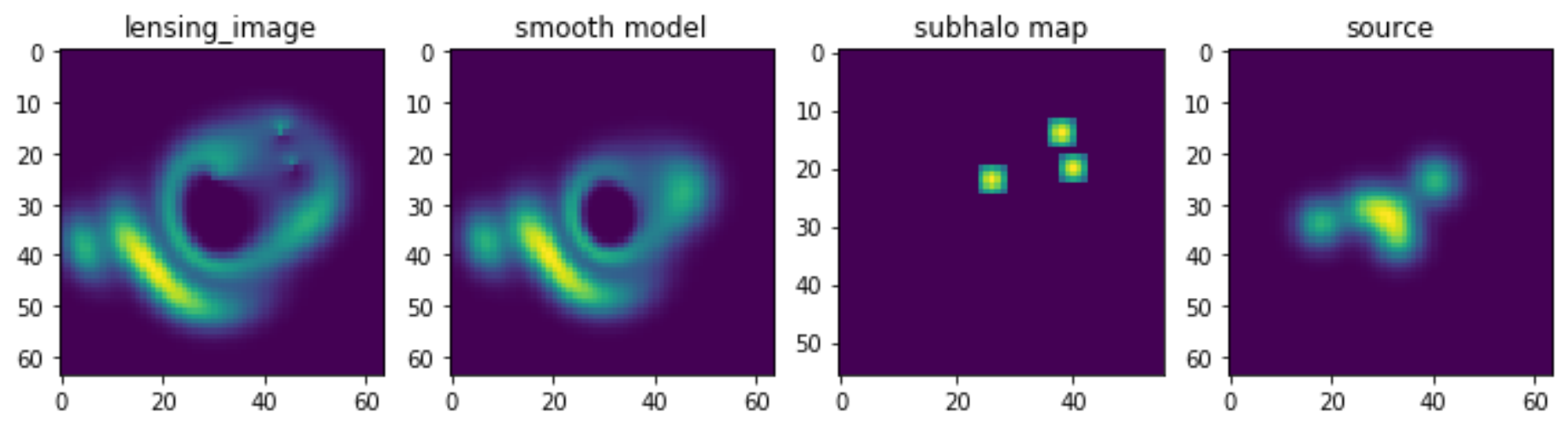}
    \caption{Simulation of Strong lensing with multiple subhalos. The figure from the left are lensing images, the smooth model, source galaxy, subhalo target map. The effect of subhalos have been magnified so that they are visible for demo.}
    \label{fig:Simulation_II}
\end{figure}

\section{Neural Network Architecture and Training Strategy}

\begin{figure}[h]
    \centering
    \includegraphics[width=0.95\columnwidth]{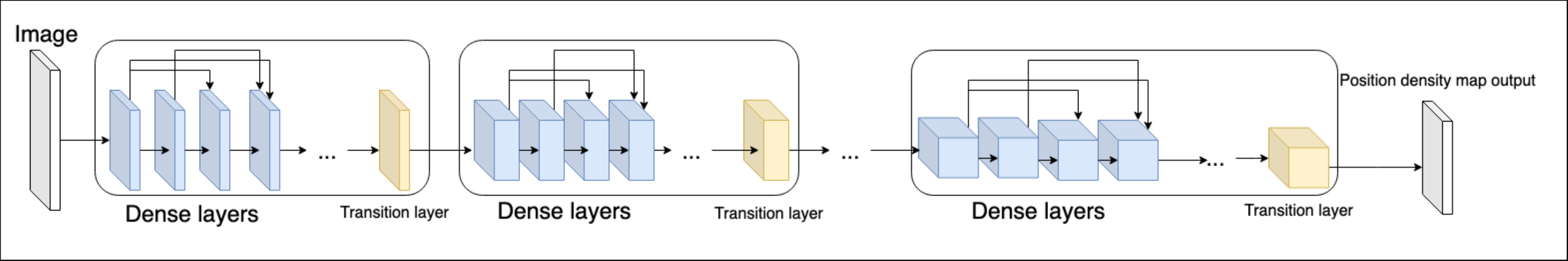}
    \caption{Schematic visualization of DenseNet}
    \label{fig:DenseNet_horizontal}
\end{figure}

We apply two states of the art CNN architecture \cite{huang2017densely} in this work:  DenseNet, and ResNet. We further modified the last few layers from the original network to make the neural network output the probability map. In order to output the probability density map, we make sure the modification of the neural network would not suffer from \emph{information bottleneck} \cite{tishby2015deep}.

A simulated lens image has a dimension of 224$\times$224. Each simulated image duplicate 3 times to make it filled in the original R,G,B channel of ResNet and DenseNet. The input images are also normalized to [0,1] before feed into the neural net. Each neural net has outputs which are both 56$\times$56$\times$1 as subhalo probability map. The output is trained against targets with Binary Cross Entropy Loss (BCELoss) of the dark matter subhalos probability map.

\section{Result}

\begin{figure}[h]
    \centering
    \includegraphics[width=0.9\columnwidth]{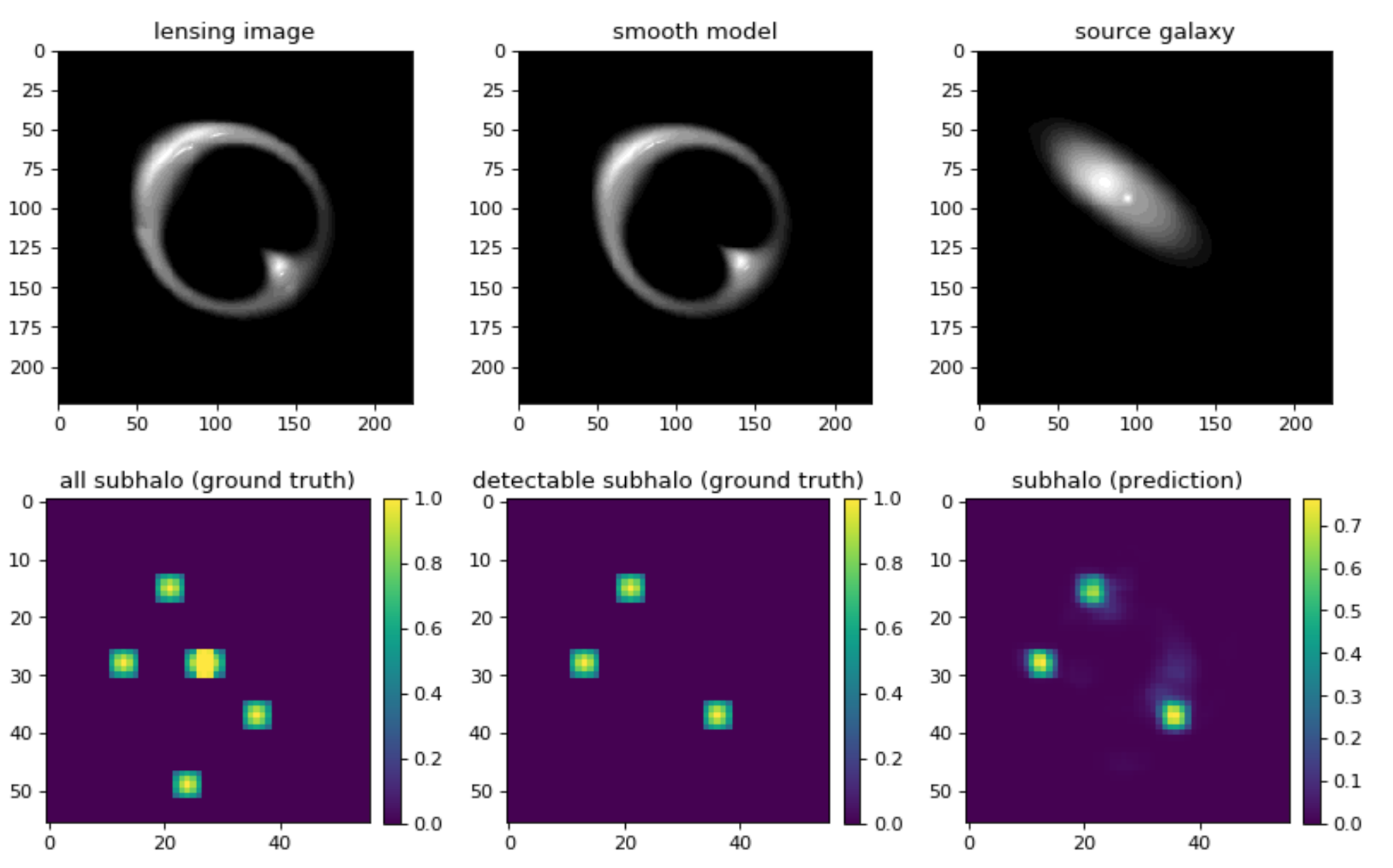}
    \caption{Lensing image and subhalos probability prediction. Noticed that there exist multiple subhalos both on the lensing arc and at the center of the lens. Only the subhalos on/near the lensing arc are treated as detectable subhalos. The neural networks predict the subhalo position correctly. }
    \label{fig:Evil_demo}
\end{figure}

For multiple subhalos, the neural networks are also able to recover the location of three individual subhalos on the probability map at Figure. \ref{fig:Evil_demo}.

% There are cases that some of the subhalos are undetectable since the subhalos are far away from the lensing arc so the perturbative effect is too small to be detected, such as Figure. 

\begin{figure}[h]
    \centering
    \includegraphics[width=1.0\columnwidth]{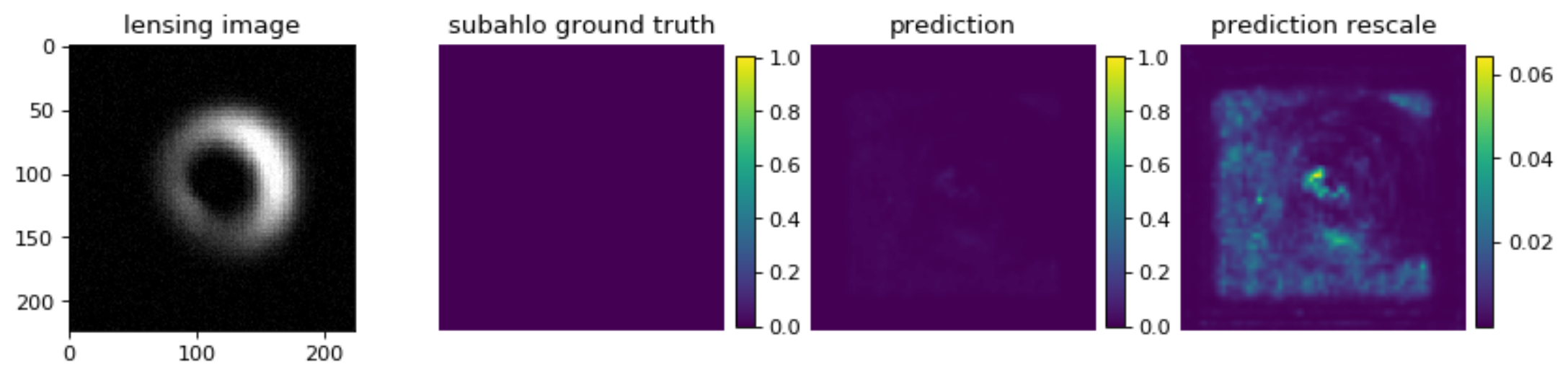}
    \caption{Lensing image and subhalos probability prediction. In this example, we have no subhalo in the system and the neural networks predict no subhalo. Interestingly, the neural network predicted lower probability where the lens is smooth and hence provide a way to reject the probability of subhalos.}
    \label{fig:serendipity_1}
\end{figure}

When there is no subhalo in the system, the probability map shows that there would be no detection. Surprisingly, when we zoom in the lower probability region, the neural networks shows that the probabilities around the lensing arc are smaller then the surround region. This shows that even though the neural networks was original designed to \emph{detect} dark matter subhalos, it actually surpass our expectation and \emph{learn to reject} the subhalos on the lensing arc of a smooth lens where there is no subhalo as shown in Figure. \ref{fig:serendipity_1}.

\section{Summary and Discussion}

In this work, we show that the neural networks are able to perform fast and automatic detection of dark matter subhalos in strong gravitational systems. We found that with using dark matter subhalos probability map as training target for the neural network, the neural networks are able to detect multiple subhalos in strong lensing image at once without any lens modeling. Furthermore, just by training the neural networks to \emph{detect} dark matter subhalos, the neural networks actually \emph{learn to reject} the subhalos on the lensing arc of a smooth lens where there is no subhalo.

Our method shows dramatically speed up in the task. It took less than a second to make a prediction on a lensing image with one graphic processing unit (GPU) once the neural networks have been trained. Traditionally, it would take more than days to for human expert to do preform precision lens modeling and further detect dark matter subhalos in strong lensing system. 

In our simulation, there are many cases where the lensing deforms from the SIE (or pure power-law)  model by the cumulative effects of subhalos on the arcs, or there might be multiple subhalos on the lensing arc that that traditional method would be difficult to deal with. While wrapping up this project, we noticed similar work by \cite{brehmer2019mining}. We would like to point out that our work focus more on detecting \emph{individual} subhalos while \cite{brehmer2019mining} focuses more on the cumulative effective of all dark matter substructures. In general, neural networks show promising results for detecting multiple subhalos as well as locates subhalos are in complicated lens systems, which could constraint dark matter models in the future.

% \section*{References}

\small
% \bibliographystyle{numeric}
% \bibliography{bibliography4.bib}
% %\printbibliography

%\bibliographystyle{unsrt}
%\bibliographystyle{numeric}
\bibliographystyle{unsrt}
\bibliography{mybib}

\begin{thebibliography}{10}

\bibitem{de2010core}
WJG De~Blok.
\newblock The core-cusp problem.
\newblock {\em Advances in Astronomy}, 2010, 2010.

\bibitem{klypin1999missing}
Anatoly Klypin, Andrey~V Kravtsov, Octavio Valenzuela, and Francisco Prada.
\newblock Where are the missing galactic satellites?
\newblock {\em The Astrophysical Journal}, 522(1):82, 1999.

\bibitem{weinberg2015cold}
David~H Weinberg, James~S Bullock, Fabio Governato, Rachel~Kuzio de~Naray, and
  Annika~HG Peter.
\newblock Cold dark matter: controversies on small scales.
\newblock {\em Proceedings of the National Academy of Sciences},
  112(40):12249--12255, 2015.

\bibitem{hezaveh2016detection}
Yashar~D Hezaveh, Neal Dalal, Daniel~P Marrone, Yao-Yuan Mao, Warren
  Morningstar, Di~Wen, Roger~D Blandford, John~E Carlstrom, Christopher~D
  Fassnacht, Gilbert~P Holder, et~al.
\newblock Detection of lensing substructure using alma observations of the
  dusty galaxy sdp. 81.
\newblock {\em The Astrophysical Journal}, 823(1):37, 2016.

\bibitem{vegetti2012gravitational}
S~Vegetti, DJ~Lagattuta, JP~McKean, MW~Auger, CD~Fassnacht, and LVE Koopmans.
\newblock Gravitational detection of a low-mass dark satellite galaxy at
  cosmological distance.
\newblock {\em Nature}, 481(7381):341, 2012.

\bibitem{despali2018modelling}
Giulia Despali, Simona Vegetti, Simon~DM White, Carlo Giocoli, and Frank~C
  van~den Bosch.
\newblock Modelling the line-of-sight contribution in substructure lensing.
\newblock {\em Monthly Notices of the Royal Astronomical Society},
  475(4):5424--5442, 2018.

\bibitem{collett2015population}
Thomas~E Collett.
\newblock The population of galaxy--galaxy strong lenses in forthcoming optical
  imaging surveys.
\newblock {\em The Astrophysical Journal}, 811(1):20, 2015.

\bibitem{hezaveh2017fast}
Yashar~D Hezaveh, Laurence~Perreault Levasseur, and Philip~J Marshall.
\newblock Fast automated analysis of strong gravitational lenses with
  convolutional neural networks.
\newblock {\em Nature}, 548(7669):555, 2017.

\bibitem{barkana1998fast}
Rennan Barkana.
\newblock Fast calculation of a family of elliptical mass gravitational lens
  models.
\newblock {\em arXiv preprint astro-ph/9802002}, 1998.

\bibitem{huang2017densely}
Gao Huang, Zhuang Liu, Laurens Van Der~Maaten, and Kilian~Q Weinberger.
\newblock Densely connected convolutional networks.
\newblock In {\em Proceedings of the IEEE conference on computer vision and
  pattern recognition}, pages 4700--4708, 2017.

\bibitem{tishby2015deep}
Naftali Tishby and Noga Zaslavsky.
\newblock Deep learning and the information bottleneck principle.
\newblock In {\em 2015 IEEE Information Theory Workshop (ITW)}, pages 1--5.
  IEEE, 2015.

\bibitem{brehmer2019mining}
Johann Brehmer, Siddharth Mishra-Sharma, Joeri Hermans, Gilles Louppe, and Kyle
  Cranmer.
\newblock Mining for dark matter substructure: Inferring subhalo population
  properties from strong lenses with machine learning.
\newblock {\em The Astrophysical Journal}, 886(1):49, 2019.

\end{thebibliography}

\end{document}